\documentclass[10pt,conference]{IEEEtran}
\usepackage{xurl}
\usepackage{amssymb}
\usepackage{pifont}
\usepackage[hidelinks]{hyperref}
\usepackage[table]{xcolor}
\definecolor{paperBlue}{RGB}{35,83,125}
\usepackage{tikz}
\usetikzlibrary{arrows.meta}

\newcommand{\logline}[2]{\noindent{\color{#1}\ttfamily\scriptsize\setlength{\baselineskip}{11pt}#2\par}}
\newcounter{logexample}
\newcommand{\logcase}[4]{%
  \refstepcounter{logexample}\label{#3}%
  \par\addvspace{6pt}\noindent
  \begingroup
  \setlength{\fboxsep}{0pt}%
  \setlength{\fboxrule}{.4pt}%
  \fcolorbox{paperBlue!35}{white}{%
    \begin{minipage}{\dimexpr\columnwidth-2\fboxrule\relax}
      \begingroup
      \setlength{\fboxsep}{5pt}%
      \colorbox{paperBlue}{\parbox{\dimexpr\linewidth-2\fboxsep\relax}{%
        {\color{white}\sffamily\bfseries\scriptsize Example~\thelogexample: #1}%
        \hfill{\color{white!90}\sffamily\scriptsize #2}}}%
      \endgroup\par
      \vspace{3pt}%
      \hspace*{5pt}\begin{minipage}{\dimexpr\linewidth-10pt\relax}
        \raggedright #4
      \end{minipage}\par\vspace{3pt}%
    \end{minipage}}%
  \endgroup
  \par\addvspace{6pt}
}

\begin{document}

\title{GH-AWTrace: A Large-Scale Dataset of GitHub Agentic Workflow Execution Logs}
\author{%
\begin{minipage}{\textwidth}
\centering
\begin{minipage}[t]{.32\textwidth}\centering
{Jasem Khelifi\par}
{\small\textit{\'ETS Montreal, University of Quebec}\\
Montreal, Canada\\
jasem.khelifi.1@ens.etsmtl.ca\par}
\end{minipage}\hfill%
\begin{minipage}[t]{.32\textwidth}\centering
{Hao Li\par}
{\small\textit{Queen's University}\\
Kingston, Canada\\
hao.li@queensu.ca\par}
\end{minipage}\hfill%
\begin{minipage}[t]{.32\textwidth}\centering
{Ali Ouni\par}
{\small\textit{\'ETS Montreal, University of Quebec}\\
Montreal, Canada\\
Ali.Ouni@etsmtl.ca\par}
\end{minipage}
\par\vspace{.8em}
\begin{minipage}[t]{.32\textwidth}\centering
{Mohammed Sayagh\par}
{\small\textit{\'ETS Montreal, University of Quebec}\\
Montreal, Canada\\
Mohammed.Sayagh@etsmtl.ca\par}
\end{minipage}\hspace{.06\textwidth}%
\begin{minipage}[t]{.32\textwidth}\centering
{Mohamed Aymen Saied\par}
{\small\textit{Laval university}\\
Quebec city, Canada\\
mohamed-aymen.saied@ift.ulaval.ca\par}
\end{minipage}
\end{minipage}%
}
\maketitle
\bstctlcite{IEEEexample:BSTcontrol}

\begin{abstract}
Agentic workflows enable AI agents to perform recurring repository tasks on schedules or in response to repository events. Existing studies examine workflow definitions and outputs, but provide limited evidence of what happens during execution. We present GH-AWTrace, a dataset linking 25,014 verified GitHub Agentic Workflows across 5,051 repositories to 4,112,526 run records and 562,877 nonempty execution log archives. The dataset preserves execution evidence subject to GitHub’s log-retention limits and links it to repository metadata and generated workflow configurations. Illustrative cases show how the logs expose tool denials, provider failures, changes in an agent’s approach, and delegation among agents. These observations enable research on execution failures, tool use, and the relationship between reported run outcomes and visible task progress. Coverage varies across workflows, and the logs provide partial observations rather than complete agent trajectories. \textbf{Video:} \url{https://youtu.be/KIlfmZhHxtA}.
\end{abstract}

\section{Introduction}

Agentic AI can inspect repositories, edit code, run tests, and resolve software issues through development tools \cite{yangSWEAgent}. In common coding-agent use, a developer starts each task by assigning an issue or writing a prompt, then reviews the proposed change \cite{githubKickOffAgent}. Agentic workflows let developers specify recurring tasks once and run them on schedules or repository events, without a person starting each execution \cite{githubAgenticWorkflows}.

GitHub Agentic Workflows (gh-aw) implements this model in Markdown: a file combines the agent's task with its trigger, permissions, and allowed outputs \cite{githubAgenticWorkflows}. gh-aw compiles the file into a GitHub Actions YAML workflow that starts a coding agent and can produce reviewable issues, comments, or pull requests \cite{githubAgenticWorkflows}. In \texttt{Azure/azure-dev}, a weekly workflow reads feedback on merged pull requests and turns recurring reviewer requests into proposed updates to the repository's Copilot instructions \cite{azureDevReviewWorkflow}. In August 2026, it examined 49 pull requests and proposed rules to reject ignored command-line flags and redact credentials from URLs before printing them in logs \cite{azureDevReviewIssue}. Repository settings prevented the workflow from opening a pull request, so it saved a branch and an issue with the patch; a maintainer refined and merged the changes through pull request \#9731 \cite{azureDevReviewIssue,azureDevReviewPR}.

Khelifi et al.~\cite{khelifi2026} studied how developers specify and maintain gh-aw workflows. Valenzuela-Toledo et al.~\cite{ghawh} released histories that link Markdown definitions to generated YAML. Sun et al.~\cite{sunCodeReview} measured whether comments from AI code-review Actions led to code changes. Wang et al.~\cite{wangAWI} used static analysis to find prompt-injection paths in agentic GitHub Actions workflows. Isbarov et al.~\cite{gitInject} tested prompt-injection attacks in live AI-powered CI/CD workflows. 
These studies provide complementary perspectives on workflow specifications, outputs, and security. However, workflow definitions describe intended behavior, while run-level outcomes summarize execution without explaining the intervening actions. Execution logs provide additional evidence of tool interactions, access restrictions, infrastructure failures, and recovery attempts. Preserving and linking this evidence is particularly important because GitHub’s retention limits make it unavailable over time.

To support research on agentic workflow execution, we release GH-AWTrace, linking 25,014 verified gh-aw workflows from 5,051 repositories to 4,112,526 run records and 562,877 nonempty execution log archives. Its contributions are threefold: (1) an archive of execution logs that may otherwise become unavailable under GitHub’s retention policies; (2) linked workflow, repository, run, and log metadata that support traceable analyses; and (3) illustrative use cases showing how the data can support studies of failures, tool interactions, and within-run agent coordination.

\section{Dataset Collection and Curation}
We construct GH-AWTrace in three stages: 1) identifying generated gh-aw workflows, 2) collecting their run metadata, and 3) archiving available execution logs. Figure \ref{fig:log-coverage} summarizes coverage across these stages.

\subsection{GitHub Agentic Workflow Identification}
First, we search for gh-aw generated workflows through the GitHub REST API
code search \cite{githubCodeSearch} using the following search query on both \texttt{.yml} and \texttt{.yaml} files separately:
\begin{center}
\setlength{\fboxsep}{5pt}
\fcolorbox{black!25}{black!3}{%
\begin{minipage}{\dimexpr\columnwidth-2\fboxsep-2\fboxrule\relax}
\ttfamily\scriptsize
"This file was automatically generated by gh-aw"\\
path:.github/workflows extension:yml fork:true\\
size:0..393216
\end{minipage}}
\end{center}
GitHub returns at most 1,000 code-search results per query \cite{githubCodeSearch}. To reach files beyond this limit, we search non-overlapping file-size ranges in bytes. When a range reported the maximum allowed 1,000 matches, we split it in half and searched both smaller ranges. We retrieve every page from the final ranges and remove duplicate files using the repository ID and exact file path.

Upon further investigation, we notice that the target phrase could appear as a quoted text rather than a generator comment. We fetched each file at the blob revision returned by search and accepted it only when the generator phrase appeared in an opening YAML comment before any configuration key. Of 25,227 candidate files in 5,059 repositories, 63 lacked the required header, and 150 indexed blobs could not be retrieved. This process yielded 25,014 verified gh-aw workflow files across 5,051 repositories.

\subsection{GitHub Agentic Workflow Runs Collection}
We use the GitHub REST API to collect runs for each gh-aw workflow, querying by workflow ID and saving all data fields returned by the run listing API \cite{githubWorkflowRuns}. We link runs to the YAML revision indexed during discovery by repository ID and exact path. This revision may differ from the one used by a run if the workflow changed. The run metadata retains its \texttt{head\_sha}. This phase yields 4,112,526 run records dated August 14, 2025, through September 21, 2026, since GitHub retention policy deletes runs after 400 days~\cite{githubRetention}.

\subsection{Log Collection}
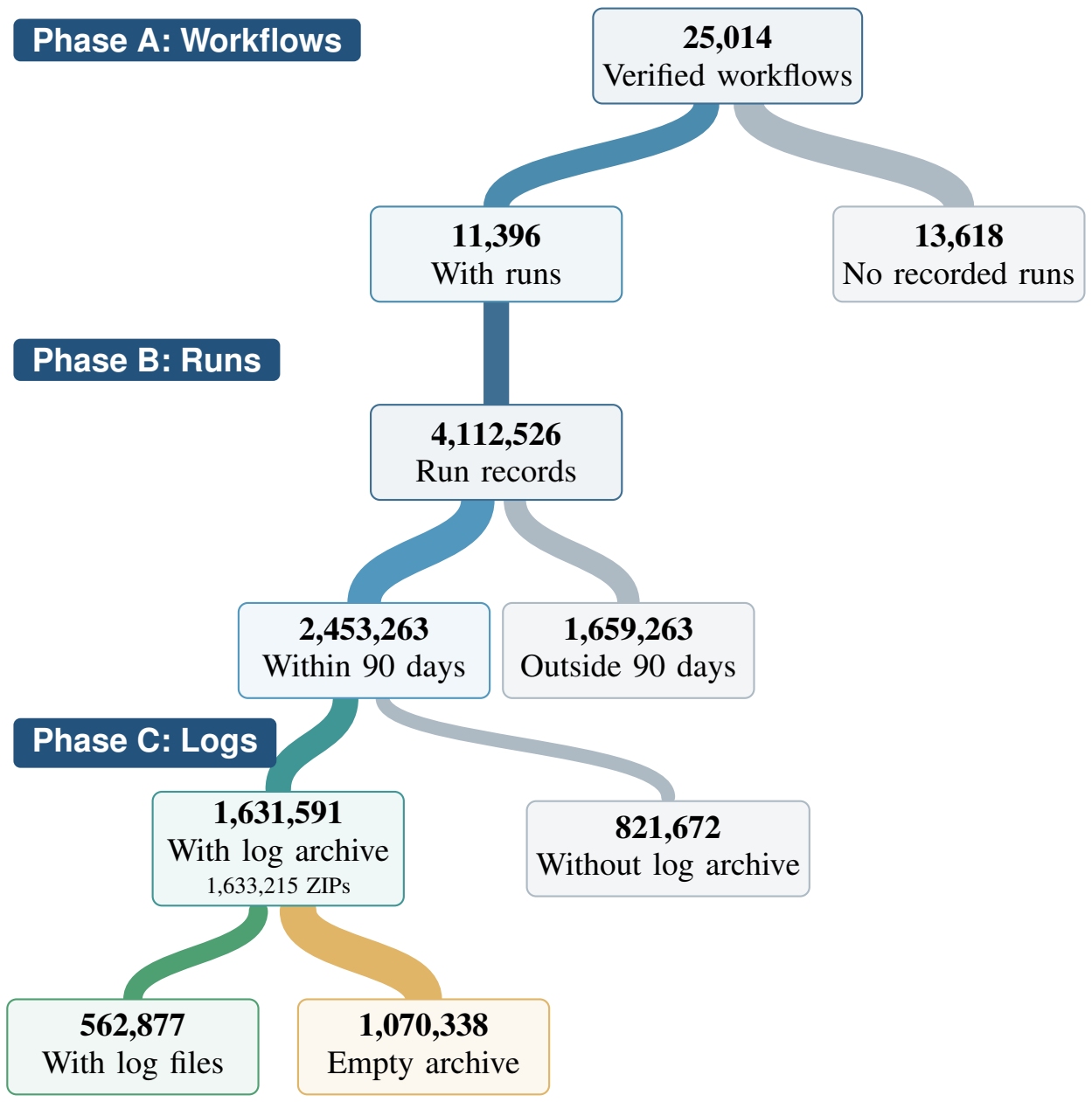
\begin{figure}[t]
\centering
\begingroup
\definecolor{flowAll}{RGB}{62,110,145}
\definecolor{flowActive}{RGB}{74,139,174}
\definecolor{flowInactive}{RGB}{173,186,196}
\definecolor{flowWindow}{RGB}{82,151,190}
\definecolor{flowOlder}{RGB}{173,186,196}
\definecolor{flowSaved}{RGB}{64,150,149}
\definecolor{flowUnavailable}{RGB}{173,186,196}
\definecolor{flowEmpty}{RGB}{224,181,103}
\definecolor{flowNonempty}{RGB}{78,159,113}
\begin{tikzpicture}[x=1cm,y=1cm,
  flow node/.style={draw=black!25,fill=white,rounded corners=2pt,
    minimum width=1.9cm,minimum height=.72cm,inner sep=2pt,
    align=center,font=\scriptsize}]
\draw[flowActive,line width=5.5pt,line cap=round]
  (5.09,-.36) .. controls (5.09,-.76) and (3.5,-.76) .. (3.5,-1.14);
\draw[flowInactive,line width=6.5pt,line cap=round]
  (5.41,-.36) .. controls (5.41,-.76) and (7,-.76) .. (7,-1.14);
\draw[flowAll,line width=5.5pt,line cap=round]
  (3.5,-1.86) -- (3.5,-2.64);
\draw[flowWindow,line width=7.2pt,line cap=round]
  (3.36,-3.36) .. controls (3.36,-3.76) and (2.5,-3.76) .. (2.5,-4.14);
\draw[flowOlder,line width=4.8pt,line cap=round]
  (3.64,-3.36) .. controls (3.64,-3.76) and (4.5,-3.76) .. (4.5,-4.14);
\draw[flowSaved,line width=5.5pt,line cap=round]
  (2.36,-4.86) .. controls (2.36,-5.21) and (1.85,-5.21) .. (1.85,-5.55);
\draw[flowUnavailable,line width=3pt,line cap=round]
  (2.64,-4.86) .. controls (2.64,-5.21) and (4.8,-5.21) .. (4.8,-5.61);
\draw[flowNonempty,line width=4.1pt,line cap=round]
  (1.7,-6.47) .. controls (1.7,-6.81) and (.75,-6.81) .. (.75,-7.14);
\draw[flowEmpty,line width=7.9pt,line cap=round]
  (2,-6.47) .. controls (2,-6.81) and (2.95,-6.81) .. (2.95,-7.14);

\node[anchor=west,fill=paperBlue,text=white,rounded corners=1.5pt,
  inner xsep=4pt,inner ysep=2pt,font=\sffamily\bfseries\scriptsize]
  at (-.16,.12) {Phase A: Workflows};
\node[anchor=west,fill=paperBlue,text=white,rounded corners=1.5pt,
  inner xsep=4pt,inner ysep=2pt,font=\sffamily\bfseries\scriptsize]
  at (-.16,-2.30) {Phase B: Runs};
\node[anchor=west,fill=paperBlue,text=white,rounded corners=1.5pt,
  inner xsep=4pt,inner ysep=2pt,font=\sffamily\bfseries\scriptsize]
  at (-.16,-5.20) {Phase C: Logs};

\node[flow node,draw=flowAll,fill=flowAll!8] at (5.25,0)
  {\textbf{25,014}\\Verified workflows};
\node[flow node,draw=flowActive,fill=flowActive!8] at (3.5,-1.5)
  {\textbf{11,396}\\With runs};
\node[flow node,draw=flowInactive,fill=flowInactive!14] at (7,-1.5)
  {\textbf{13,618}\\No recorded runs};
\node[flow node,draw=flowAll,fill=flowAll!8] at (3.5,-3)
  {\textbf{4,112,526}\\Run records};
\node[flow node,draw=flowWindow,fill=flowWindow!8] at (2.5,-4.5)
  {\textbf{2,453,263}\\Within 90 days};
\node[flow node,draw=flowOlder,fill=flowOlder!14] at (4.5,-4.5)
  {\textbf{1,659,263}\\Outside 90 days};
\node[flow node,draw=flowSaved,fill=flowSaved!8] at (1.85,-6)
  {\textbf{1,631,591}\\With log archive\\[-1pt]{\tiny 1,633,215 ZIPs}};
\node[flow node,draw=flowUnavailable,fill=flowUnavailable!14] at (4.8,-6)
  {\textbf{821,672}\\Without log archive};
\node[flow node,draw=flowNonempty,fill=flowNonempty!8] at (.75,-7.5)
  {\textbf{562,877}\\With log files};
\node[flow node,draw=flowEmpty,fill=flowEmpty!10] at (2.95,-7.5)
  {\textbf{1,070,338}\\Empty archive};
\end{tikzpicture}
\endgroup
\caption{Workflow, run, and log coverage across the three collection phases. The final split counts attempt ZIPs, including reruns.}
\label{fig:log-coverage}
\end{figure}

Finally, GitHub automatically deletes public workflow execution logs after at most 90 days \cite{githubRetention}. We therefore select 2,453,263 run records created from June 23 through September 21, 2026, the 90 days before our run cutoff. For these runs, we download and save 1,633,215 execution log archives (56.1~GB), one ZIP per downloaded attempt \cite{githubWorkflowRuns}. A run has one GitHub run ID. Each rerun keeps that ID and receives a new attempt number, so we identify an archive by run ID and attempt number. We check each ZIP's integrity and record its SHA-256 hash and file list. Of the selected runs, 821,672 have no downloaded archive, which may reflect retention, deletion, access restrictions, or an unfinished run. Among the downloaded ZIPs, 1,070,338 contain no files and 562,877 contain log files. The 562,877 archives with files belong to 561,367 runs, and 18,337 archives come from attempts after the first. In an early audit of 2,900 empty ZIPs with saved job metadata, 2,699 show only skipped jobs in the available pages and 195 list no jobs. We cannot determine the cause of every empty ZIP. Figure~\ref{fig:log-coverage} summarizes workflow, run-record, and log-archive coverage.

\section{Dataset Characteristics}

\subsection{Schema and Stored Files}
\begin{figure*}[t]
\centering
\begingroup
\colorlet{schemaAccent}{paperBlue}
\begin{tikzpicture}[x=1cm,y=.88cm]
\newcommand{\entitycard}[2]{%
  \filldraw[rounded corners=2pt,fill=black!4,draw=black!16,line width=.45pt]
    (#1,0) rectangle ({#1+3.75},-5.17);
  \fill[schemaAccent,rounded corners=2pt] (#1,0) rectangle ({#1+3.75},-.54);
  \fill[schemaAccent] (#1,-.37) rectangle ({#1+3.75},-.54);
  \node[anchor=west,font=\bfseries\small,text=white] at ({#1+.19},-.27) {#2};
}
\newcommand{\entityfield}[5]{%
  \ifnum#5=1
    \fill[schemaAccent] ({#1+.18},#2) circle[radius=.045];
  \fi
  \node[anchor=west,font=\ttfamily\scriptsize,text=black!85]
    at ({#1+.33},#2) {#3};
  \node[anchor=east,font=\scriptsize,text=black!55]
    at ({#1+3.58},#2) {#4};
}
\newcommand{\entityfile}[2]{%
  \fill[schemaAccent!8] ({#1+.08},-4.67) rectangle ({#1+3.67},-5.08);
  \node[anchor=west,font=\ttfamily\tiny,text=schemaAccent!85]
    at ({#1+.20},-4.88) {#2};
}
\entitycard{0}{Repository}
\entitycard{4.55}{Workflow}
\entitycard{9.10}{Run}
\entitycard{13.65}{Attempt log}

\entityfield{0}{-.91}{id}{integer}{1}
\entityfield{0}{-1.33}{full\_name}{text}{0}
\entityfield{0}{-1.75}{stars}{integer}{0}
\entityfield{0}{-2.17}{fetched\_at}{time}{0}
\entityfield{0}{-2.59}{metadata}{JSON}{0}
\entityfile{0}{repository.json}

\entityfield{4.55}{-.91}{repo\_id}{integer}{1}
\entityfield{4.55}{-1.33}{path}{text}{1}
\entityfield{4.55}{-1.75}{workflow\_id}{integer}{0}
\entityfield{4.55}{-2.17}{name}{text}{0}
\entityfield{4.55}{-2.59}{sha}{text}{0}
\entityfield{4.55}{-3.01}{verified}{boolean}{0}
\entityfield{4.55}{-3.43}{state}{text}{0}
\entityfield{4.55}{-3.85}{metadata}{JSON}{0}
\entityfield{4.55}{-4.27}{discovered\_at}{time}{0}
\entityfile{4.55}{source .yml / .yaml}

\entityfield{9.10}{-.91}{id}{integer}{1}
\entityfield{9.10}{-1.33}{repo\_id}{integer}{1}
\entityfield{9.10}{-1.75}{path}{text}{1}
\entityfield{9.10}{-2.17}{created\_at}{time}{0}
\entityfield{9.10}{-2.59}{status}{text}{0}
\entityfield{9.10}{-3.01}{conclusion}{text}{0}
\entityfield{9.10}{-3.43}{attempts}{integer}{0}
\entityfield{9.10}{-3.85}{verification}{text}{0}
\entityfield{9.10}{-4.27}{metadata}{JSON}{0}
\entityfile{9.10}{run-page JSON / run.json}

\entityfield{13.65}{-.91}{run\_id}{integer}{1}
\entityfield{13.65}{-1.33}{attempt}{integer}{1}
\entityfield{13.65}{-1.75}{kind}{text}{1}
\entityfield{13.65}{-2.17}{state}{text}{0}
\entityfield{13.65}{-2.59}{path}{text}{0}
\entityfield{13.65}{-3.01}{bytes}{integer}{0}
\entityfield{13.65}{-3.43}{sha256}{text}{0}
\entityfield{13.65}{-3.85}{files}{integer}{0}
\entityfield{13.65}{-4.27}{error}{text}{0}
\entityfile{13.65}{logs.zip / logs.receipt.json}

\draw[schemaAccent,line width=.7pt,-{Latex[length=1.4mm]}]
  (3.77,-.91) -- (4.53,-.91);
\node[font=\tiny,text=schemaAccent,fill=white,inner sep=.6pt]
  at (4.15,-.67) {1:N};
\draw[schemaAccent,line width=.7pt,-{Latex[length=1.4mm]}]
  (8.32,-1.33) -- (8.70,-1.33) -- (8.70,-1.75) -- (9.08,-1.75);
\node[font=\tiny,text=schemaAccent,fill=white,inner sep=.6pt]
  at (8.70,-.99) {1:N};
\draw[schemaAccent,line width=.7pt,-{Latex[length=1.4mm]}]
  (12.87,-.91) -- (13.63,-.91);
\node[font=\tiny,text=schemaAccent,fill=white,inner sep=.6pt]
  at (13.25,-.67) {1:N};
\end{tikzpicture}
\endgroup
\caption{Dataset schema. Colored dots mark key and join fields. Connectors show one-to-many relationships, and shaded footers name preserved files.}
\label{fig:data-model}
\end{figure*}
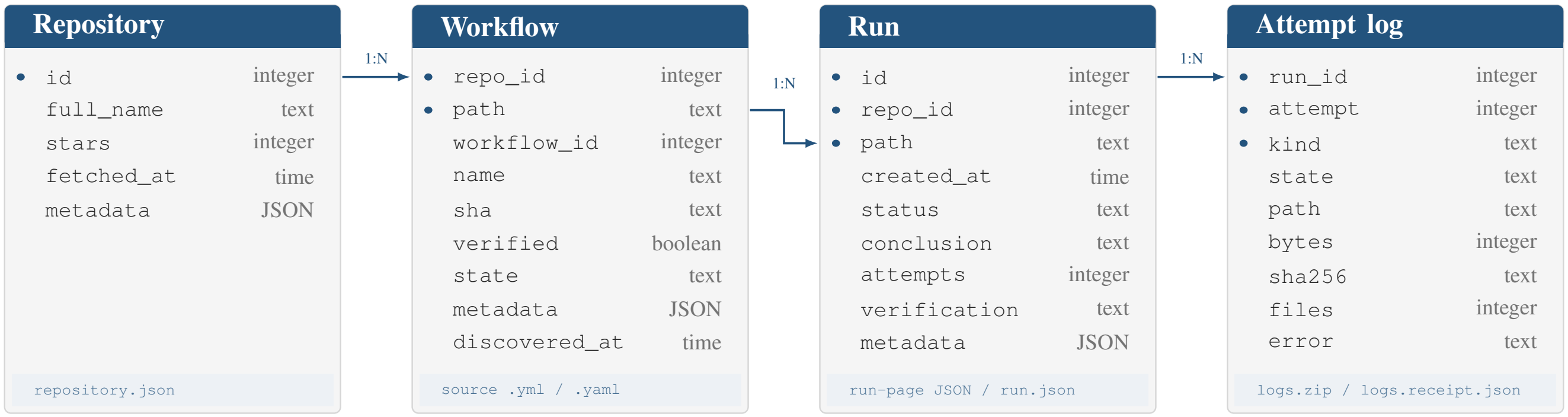

Figure~\ref{fig:data-model} shows how repositories, gh-aw workflows, runs, and log archives are linked. A repository can contain one or more gh-aw workflows, linked through \texttt{repo\_id}. When a workflow executes, it has one or more runs, linked through \texttt{repo\_id} and its exact YAML \texttt{path}. A run may have no downloaded log archive or an archive linked through \texttt{run\_id}. If a run is rerun, each downloaded attempt archive is distinguished by its \texttt{attempt} number.

\subsection{Available Content}
The JSON fields retain additional GitHub metadata, including each run's event, actors, branch, commit SHA, and timestamps. CSV indexes support direct joins, while the raw YAML, API responses, and ZIPs preserve the underlying content. A download row records whether an archive was saved or unavailable; a saved ZIP can still contain zero files. Its receipt lists each member and checksum. Runs skipped because of age have no downloaded archive.

\subsection{Execution Distributions}
\begin{figure}[t]
\centering
\begin{tikzpicture}[x=1cm,y=.70cm]
\scriptsize
\definecolor{violinFill}{RGB}{175,217,241}
\definecolor{violinGrid}{RGB}{224,234,242}
\draw[violinGrid,line width=.35pt] (1.10,0.000) -- (6.12,0.000); \node[anchor=east,text=paperBlue] at (.98,0.000) {1};
\draw[violinGrid,line width=.35pt] (1.10,1.000) -- (6.12,1.000); \node[anchor=east,text=paperBlue] at (.98,1.000) {10};
\draw[violinGrid,line width=.35pt] (1.10,2.000) -- (6.12,2.000); \node[anchor=east,text=paperBlue] at (.98,2.000) {100};
\draw[violinGrid,line width=.35pt] (1.10,3.000) -- (6.12,3.000); \node[anchor=east,text=paperBlue] at (.98,3.000) {1,000};
\draw[violinGrid,line width=.35pt] (1.10,4.000) -- (6.12,4.000); \node[anchor=east,text=paperBlue] at (.98,4.000) {10,000};
\draw[paperBlue,line width=.6pt] (1.10,0) -- (1.10,4.826);
\draw[paperBlue,line width=.6pt] (1.10,0) -- (6.12,0);
\node[rotate=90,text=paperBlue] at (.15,2.413) {Logs per workflow};
\filldraw[fill=violinFill,draw=paperBlue,line width=.8pt] plot[smooth] coordinates {
(3.247,0.000) (3.168,0.193) (3.109,0.386) (3.003,0.579) (2.853,0.772) (2.754,0.965)
(2.771,1.158) (2.796,1.351) (2.727,1.544) (2.699,1.737) (2.874,1.931) (3.154,2.124)
(3.360,2.317) (3.465,2.510) (3.513,2.703) (3.539,2.896) (3.561,3.089) (3.583,3.282)
(3.599,3.475) (3.607,3.668) (3.611,3.861) (3.615,4.054) (3.618,4.247) (3.619,4.440)
(3.619,4.633) (3.620,4.826) (3.620,4.826) (3.621,4.633) (3.621,4.440) (3.622,4.247)
(3.625,4.054) (3.629,3.861) (3.633,3.668) (3.641,3.475) (3.657,3.282) (3.679,3.089)
(3.701,2.896) (3.727,2.703) (3.775,2.510) (3.880,2.317) (4.086,2.124) (4.366,1.931)
(4.541,1.737) (4.513,1.544) (4.444,1.351) (4.469,1.158) (4.486,0.965) (4.387,0.772)
(4.237,0.579) (4.131,0.386) (4.072,0.193) (3.993,0.000)
} -- cycle;
\draw[paperBlue,line width=2pt,line cap=round] (3.62,0.778) -- (3.62,1.778);
\filldraw[fill=white,draw=paperBlue,line width=.7pt] (3.62,1.204) circle[radius=.065];
\node[anchor=north,text=paperBlue] at (3.62,-.16) {Workflows with logs};
\end{tikzpicture}
\caption{Distribution of logs per workflow across 6,040 workflows. The log-scaled axis shows the median (dot) and interquartile range (line).}
\label{fig:logs-per-workflow}
\end{figure}
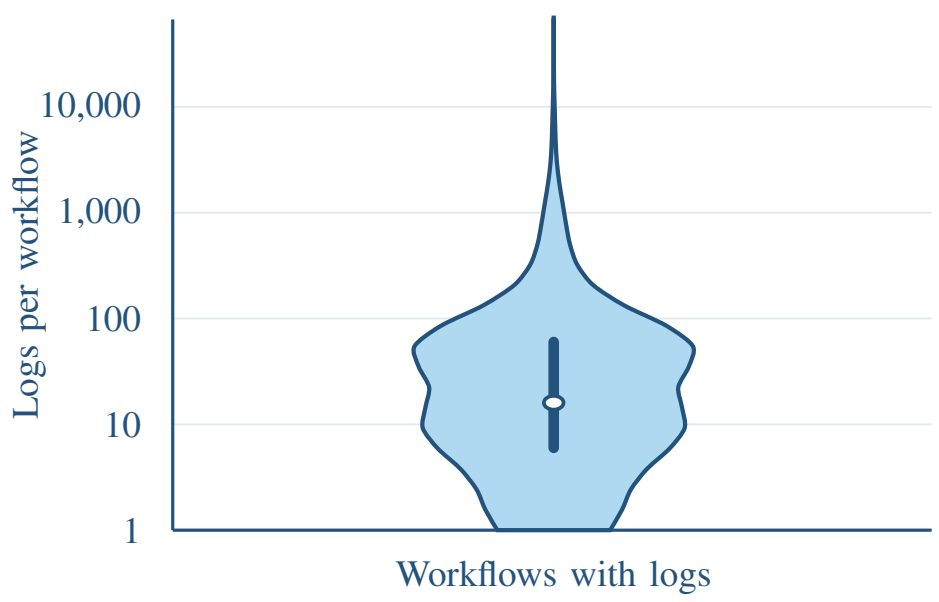

Figure~\ref{fig:logs-per-workflow} shows the distribution for the 6,040 workflows with logs. The median is 16 logs per workflow (interquartile range: 6--60). Each downloaded execution attempt contributes one log archive.

Table~\ref{tab:triggers} groups nonempty log archives by their run's recorded event. The workflow column counts verified workflows with at least one log under that event. A workflow can appear under multiple events, so workflow counts are not additive. Percentages use the 6,040 workflows with logs and 562,877 nonempty archives in the fixed 90-day window.

\begin{table}[t]
\centering
\caption{Workflows and nonempty logs by observed run trigger.}
\label{tab:triggers}
\begingroup
\definecolor{triggerStripe}{RGB}{228,239,248}
\scriptsize
\setlength{\tabcolsep}{5pt}
\renewcommand{\arraystretch}{1.08}
\rowcolors{2}{triggerStripe}{white}
\begin{tabular}{lrr}
\hline
\rowcolor{paperBlue}
\color{white}\bfseries Trigger & \color{white}\bfseries Workflows & \color{white}\bfseries Logs \\
\hline
schedule & 4,766 (78.91\%) & 252,691 (44.89\%) \\
pull\_request\_target & 59 (0.98\%) & 73,594 (13.07\%) \\
pull\_request & 365 (6.04\%) & 61,224 (10.88\%) \\
issue\_comment & 171 (2.83\%) & 58,312 (10.36\%) \\
workflow\_dispatch & 971 (16.08\%) & 46,165 (8.20\%) \\
issues & 304 (5.03\%) & 26,719 (4.75\%) \\
workflow\_run & 156 (2.58\%) & 18,944 (3.37\%) \\
check\_suite & 3 (0.05\%) & 12,061 (2.14\%) \\
pull\_request\_review\_comment & 23 (0.38\%) & 6,351 (1.13\%) \\
push & 101 (1.67\%) & 3,375 (0.60\%) \\
check\_run & 5 (0.08\%) & 2,063 (0.37\%) \\
discussion & 11 (0.18\%) & 617 (0.11\%) \\
deployment\_status & 13 (0.22\%) & 425 (0.08\%) \\
pull\_request\_review & 5 (0.08\%) & 201 (0.04\%) \\
release & 13 (0.22\%) & 56 (0.01\%) \\
discussion\_comment & 8 (0.13\%) & 46 ($<0.01$\%) \\
repository\_dispatch & 3 (0.05\%) & 21 ($<0.01$\%) \\
create & 3 (0.05\%) & 12 ($<0.01$\%) \\
\hline
\rowcolor{triggerStripe}
\bfseries Total with logs & \bfseries 6,040 (100.00\%) & \bfseries 562,877 (100.00\%) \\
\hline
\end{tabular}
\endgroup
\end{table}

\section{Analysis \& Potential Use Cases}
The excerpts below reproduce text from each cited run's archived attempt-1 agent job log. We omit repeated leading UTC timestamps. Colors are editorial, and ellipses mark omitted text or intervening lines.

\subsection{Agentic Workflow Failure Analysis}
Of 562,877 nonempty log archives, 178,151 (31.7\%) are linked to run records with a \texttt{failure} conclusion. This does not attribute every run failure to its agent. Examples~\ref{ex:issue-triage} and~\ref{ex:failure-scanner} show a blocked task and a provider failure, respectively.

\logcase{ISSUE TRIAGE}{Run 30125677266 \cite{runIssueTriage}}{ex:issue-triage}{%
  \logline{red!75!black}{\ding{55} issue\_read (MCP: github) \textperiodcentered{} method: "get", \ldots{}}
  \logline{red!75!black}{\ensuremath{\llcorner} MCP server 'github': McpError: MCP error 0: [Filtered]}
  \logline{red!75!black}{filtered by integrity policy (Resource}
  \logline{blue!70!black}{\ldots{} I cannot read its content to perform triage.}
  \logline{black!70}{\ensuremath{\bullet} report\_incomplete (MCP: safeoutputs) \textperiodcentered{} reason: \ldots{}}
}

\logcase{FAILURE SCANNER}{Run 35294030895 \cite{runFailureScanner}}{ex:failure-scanner}{%
  \logline{blue!70!black}{New, stable QUIC failure. I'll pair \ldots{}}
  \logline{red!75!black}{Authentication failed with provider at http://172.30.0.30:10002 (HTTP 403).}
  \logline{black!70}{[copilot-harness] attempt 1: provider authentication failed after partial execution - will retry once \ldots{}}
  \logline{red!75!black}{[copilot-harness] attempt 2: authentication failed --- not retrying}
  \logline{red!75!black}{\#\#[error]Process completed with exit code 1.}
}
The scanner's two attempts are retries inside one GitHub run. Such paired messages allow studies of task, access-policy, and infrastructure failures, including whether recovery was attempted.

\subsection{Visible Agent Decision Traces}
Agent-job logs can record a prompt-file path and size, visible agent messages, tool calls, and responses. In the cases below, the full prompt text is absent, so these traces should not be treated as complete prompts or private reasoning. In Examples~\ref{ex:powertoys-triage} and~\ref{ex:dependency-check}, one agent changes its publishing command after a tool denial, while another inspects a dependency update before posting its verdict.

In these Copilot runs, the generated YAML labels the execution step \texttt{Execute GitHub Copilot CLI} with \texttt{id: agentic\_execution}. Its job log prints \texttt{[INFO] Executing agent command...} before the harness starts. The ZIP member \texttt{*\_agent.txt} is a whole job log, not a role label for each message.

\logcase{ISSUE TRIAGE}{Run 35351834895 \cite{runPowerToysTriage}}{ex:powertoys-triage}{%
  \logline{black!70}{\ensuremath{\bullet} Read issue-context.md}
  \logline{red!75!black}{\ding{55} Publish triage summary for issue \#50643 (shell)}
  \logline{red!75!black}{\ensuremath{\llcorner} Permission to run this tool was denied due to the following rules: `shell(cat)`}
  \logline{blue!70!black}{Let me use a different approach:}
  \logline{green!45!black}{Perfect! The triage summary has been successfully published.}
}

\logcase{DEPENDENCY CHECK}{Run 33057082627 \cite{runRequirementsCheck}}{ex:dependency-check}{%
  \logline{blue!70!black}{Now I need to check the diff between 2.4.1 and 2.5.0 for async\_blocking and security, and check the PR link.}
  \logline{black!70}{\ensuremath{\bullet} Get diff between 2.4.1 and 2.5.0 (shell)}
  \logline{blue!70!black}{Now I have all the info needed. Let me compile the verdicts:}
  \logline{black!70}{\ensuremath{\bullet} add\_comment (MCP: safeoutputs) \textperiodcentered{} body: \ldots{}}
  \logline{green!45!black}{\ensuremath{\llcorner} {\{"result":"success","temporary\_id":"aw\_k0yXps72",\linebreak
    "comment":"\#aw\_k0yXps72"\}}}
}
These sequences support studies of tool selection, adaptation after denial, and how evidence is turned into an action. The logged task statements are agent output, not a substitute for the complete prompt.

\subsection{Multi-Agent Coordination}
Example~\ref{ex:expert-review} shows a review run that dispatched three background reviewers and used agreement between the two respondents to select findings. It shows within-run delegation and synthesis, while providing no evidence of memory shared across runs.

\logcase{EXPERT CODE REVIEW}{Run 31484611735 \cite{runExpertReview}}{ex:expert-review}{%
  \logline{blue!70!black}{[7] \ensuremath{\blacklozenge} This is a documentation-only PR (4 files, all docs/README). Now dispatching 3 parallel expert reviewers.}
  \logline{black!70}{[8] \ding{51} task(name: reviewer-1, description: Reviewer 1: deep reasoning review, \ldots{})}
  \logline{black!70}{[10] \ding{51} task(name: reviewer-3, description: Reviewer 3: alternative perspective revi\ldots{}, \ldots{})}
  \logline{black!70}{[97] \ding{51} read\_agent(agent\_id: reviewer-2, timeout: 180, wait: true)}
  \logline{green!45!black}{[101] \ensuremath{\blacklozenge} Reviewer 3 produced no response → 2-reviewer mode. Both completed reviewers agree (2/2) \ldots{}}
}
This trace supports analysis of division of labor, missing responses, and how agreement among agents changes a final review.

\section{Conclusion}



We presented GH-AWTrace, a dataset linking 25,014 verified GitHub Agentic Workflows across 5,051 repositories to 4,112,526 run records and 562,877 nonempty execution log archives. By preserving execution evidence alongside workflow and run metadata, the dataset supports analyses that cannot be conducted from workflow definitions or run conclusions alone. Illustrative cases reveal access restrictions, provider failures, changes in agent actions, and within-run delegation.

The dataset provides partial observations of execution. Log availability is constrained by retention and retrieval coverage, and nonempty archives do not necessarily contain complete agent traces. Analyses should therefore distinguish workflow runs from execution attempts, reported outcomes from observed task completion, and agent behavior from infrastructure failures. Concrete research directions include developing failure taxonomies, examining recovery after tool denials, and investigating agreement between run conclusions and visible task progress. GH-AWTrace provides preserved evidence for pursuing these questions, while broader claims require systematic sampling and validation.

\section*{Data Availability}
GH-AWTrace, including its linked data tables, nonempty log archives, schema, and access instructions, is publicly available at \url{https://huggingface.co/datasets/jaykay9999/GHAWTrace}. A \href{https://huggingface.co/datasets/jaykay9999/GHAWTrace/blob/8804e51fe41ddd6fbdc4eccb23d1734a8a6cec11/examples/reproduce.py}{Python example} joins the tables, reproduces the 31.7\% result, and extracts a cited agent log.

\bibliographystyle{IEEEtran}
\bibliography{references}

\end{document}